\documentstyle[sprocl,epsfig]{article}

\def\Journal#1#2#3#4{{#1} {\bf #2}, #3 (#4)}

\def\NPB{{\em Nucl. Phys.} B}
\def\PLB{{\em Phys. Lett.}  B}
\def\PRL{\em Phys. Rev. Lett.}
\def\PRD{{\em Phys. Rev.} D}

\def\be{\begin{equation}}
\def\ee{\end{equation}}
\def\bea{\begin{eqnarray}}
\def\eea{\end{eqnarray}}
\newcommand{\simorder}{\raisebox{-4pt}{$\, \stackrel{\textstyle >}{\sim} \,$}}

\begin{document}

\renewcommand{\thefootnote}{\fnsymbol{footnote}}

\title{TRANSVERSE \protect$\Lambda$
 POLARIZATION IN UNPOLARIZED SEMI-INCLUSIVE DIS
$\!$\footnote{Talk delivered by F. Murgia at the IX International
Workshop on Deep Inelastic Scattering (DIS2001),
Bologna, 27 April - 1 May 2001.}
\vspace*{-10pt}}

\author{M. ANSELMINO$^a$, D. BOER$^b$, U. D'ALESIO$^c$,
F. MURGIA$^c$}

\address{\vspace{2pt}
$^a$Dipartimento di Fisica Teorica, Universit\`a di Torino,
and \\ INFN, Sezione di Torino,
 Via P. Giuria 1, I-10125 Torino, Italy\\ \vspace{2pt}
$^b$RIKEN-BNL Research Center, Brookhaven National Laboratory,\\
Upton, New York 11973, USA\\ \vspace{2pt}
$^c$INFN, Sezione di Cagliari, and Dipartimento di Fisica,
Universit\`a di Cagliari,\\
C.P. 170, I-09042 Monserrato (CA), Italy}

\maketitle\abstracts{
The long-standing problem of transverse $\Lambda$ polarization in
high-energy collisions of unpolarized hadrons can be tackled by
considering new, spin and $\mbox{\boldmath$k$}_\perp$-dependent
quark fragmentation functions for an
unpolarized quark into a polarized, spin-1/2 hadron.
Simple phenomenological parameterizations of these new ``polarizing
fragmentation functions'', which describe quite well the experimental
data on $\Lambda$ and $\bar\Lambda$ hyperons produced in $p-A$ processes,
are utilized and extended here to give predictions for transverse $\Lambda$
polarization in semi-inclusive DIS.}

\section{Transverse $\Lambda$ polarization in hadronic collisions}

Transverse hyperon polarization in high-energy, unpolarized
hadron collisions is a long-standing challenge for theoretical
models of hadronic reactions.
We have recently proposed an approach~\cite{abdm01} to this problem
based on perturbative QCD and its factorization theorems, and including
polarization and intrinsic transverse momentum,
$\mbox{\boldmath$k$}_\perp$, effects.
This approach was already applied to the study of transverse
single spin asymmetries in inclusive particle production at large $x_{_F}$
and medium-large $p_{_T}$.~\cite{abm95}
It requires the introduction of a new
class of leading-twist, polarized and
$\mbox{\boldmath$k$}_\perp$-dependent distributions and
fragmentation functions (FF).
These new functions can be extracted
by fitting available experimental data
and consistently applied to give predictions for other processes.

A large amount of data on transverse $\Lambda$ polarization,
$P_{_T}^\Lambda$, in unpolarized hadronic collisions is available;
the main properties of experimental data at $x_{_F}\simorder0.2$
can be summarized as follows: 1)~$P_{_T}^\Lambda < 0$;
2)~Starting from zero at very low $p_{_T}$, $|P_{_T}^\Lambda|$ increases
up to $p_{_T}\sim 1$ GeV, where it flattens
to an almost constant value, up to the highest measured $p_{_T}$
of about $3$ GeV;
3)~The value of $|P_{_T}^\Lambda|$ in this plateau regime increases almost
linearly with $x_{_F}$; 4)~$P_{_T}^{\bar\Lambda}$ is compatible
with zero.
In our approach, the transverse hyperon
polarization in unpolarized hadro\-nic reactions at large
$p_{_T}$ can be written,
e.g. for the $p p\to\Lambda^\uparrow\,X$ case , as follows~\cite{abdm01}

\vspace{-6pt}

\begin{eqnarray}
 &\,\,\,\,P_{_T}^{\Lambda}(x_{_F},p_{_T})&\, = \,
  \frac{d\sigma^{pp\to\,\Lambda^\uparrow\,X}-
   d\sigma^{pp\to\,\Lambda^\downarrow\,X}}
  {d\sigma^{pp\to\,\Lambda^\uparrow\,X}+
   d\sigma^{pp\to\,\Lambda^\downarrow\,X}} \label{ptlh}\\
 &\!\!\!\!\!\!\!\!\!\!\!\!\!\!\!\!\!\!\!\!\!\!\!\!=\!\!\!\!\!\!&
  \!\!\!\!\!\!\!\!\!\!\!\!\!\!\!\!\!\!\!\!\!
 \frac{\sum\,\int dx_a\,dx_b \int\! d^2\mbox{\boldmath $k$}_{\perp c}\,
 f_{a/p}(x_a)\, f_{b/p}(x_b)\, d\hat\sigma(x_a,x_b;
 \mbox{\boldmath $k$}_{\perp c})\,
 \Delta^{\!N}\!D_{\Lambda^\uparrow\!/c}(z,\mbox{\boldmath $k$}_{\perp c})}
 {\sum\,\int dx_a\,dx_b \int\! d^2\mbox{\boldmath $k$}_{\perp c}\,
 f_{a/p}(x_a)\, f_{b/p}(x_b)\, d\hat\sigma(x_a,x_b;
 \mbox{\boldmath $k$}_{\perp c})\,
 D_{\Lambda/c}(z,\mbox{\boldmath $k$}_{\perp c})}\,,\nonumber
\end{eqnarray}

\noindent
where $d\sigma^{pp\to\Lambda\,X}$ stands for $E_{\Lambda}
d\sigma^{pp\to\Lambda\,X}/d^3\mbox{\boldmath $p$}_\Lambda$;
$f_{a/p}(x_a)$ are the usual unpolarized parton densities;
$d\hat\sigma(x_a,x_b;\mbox{\boldmath $k$}_{\perp c})$ is the
lowest order partonic cross section with the inclusion of
$\mbox{\boldmath $k$}_{\perp c}$ effects; 
$D_{\Lambda/c}(z,\mbox{\boldmath $k$}_{\perp c})$ and
$\Delta^{\!N}\!D_{\Lambda^\uparrow\!/c}(z,\mbox{\boldmath $k$}_{\perp c})$
are respectively the unpolarized and the {\it polarizing} 
FF \cite{abdm01,muta96} for the process $c\to\Lambda+X$.

Eq.~(\ref{ptlh})
is based on some simplifying conditions:
1)~As suggested by experimental data, the
$\Lambda$ polarization is assumed to be generated in the
fragmentation process; 2)~$\Lambda$ FF include also
$\Lambda$'s coming from decays of other hyperon resonances.
In order to reduce the number of parameters, 
as a first step the full integration over
$\mbox{\boldmath$k$}_{\perp c}$ is replaced by evaluation at
an {\it effective}, average $\langle k_\perp^0(z)\rangle$;
 $\langle k_\perp^0(z)\rangle$ and
$\Delta^{\!N}\!D_{\Lambda^\uparrow\!/c}(z,\langle k_\perp^0\rangle)$
are then parameterized by using simple expressions of the form
$Nz^a(1-z)^b$.
We impose appropriate positivity bounds on
$\Delta^{\!N}\!D_{\Lambda^\uparrow\!/c}$
and consider only leading (or valence, $q_v$) quarks
in the fragmentation process.
In this way, a very good fit to experimental data for $\Lambda$ and
$\bar\Lambda$ polarization, at $p_{_T} \simorder 1$ GeV,
can be obtained.~\cite{abdm01} Moreover, it results that
$\Delta^{\!N}\!D_{\Lambda^\uparrow\!/{u,d}}<0$,
$\Delta^{\!N}\!D_{\Lambda^\uparrow\!/{s}}>0$, and
$\Delta^{\!N}\!D_{\Lambda^\uparrow\!/{s}}>
|\Delta^{\!N}\!D_{\Lambda^\uparrow\!/{u,d}}|$.
Notice that these general features are similar to
those expected for the longitudinally polarized FF,
$\Delta D_{\Lambda/q}(z)$, in the well-known
Burkardt-Jaffe model.~\cite{buja93}

\section{$P_{_T}^\Lambda$ and $P_{_T}^{\bar\Lambda}$
in semi-inclusive DIS at $x_{_F}>0$}

We want now to extend our analysis to the case of $\Lambda$ polarization
in unpolarized semi-inclusive DIS (SIDIS),
$\ell p\to \ell'\Lambda^\uparrow\, X$.
We neglect the intrinsic $\mbox{\boldmath$k$}_\perp$ effects in
the unpolarized initial proton. Then, at leading twist and leading order,
in the virtual boson-proton c.m. reference frame the virtual boson-quark
scattering is collinear, and the intrinsic transverse momentum of the
$\Lambda$ with respect to the fragmenting quark and its observed
transverse momentum $\mbox{\boldmath$p$}_T$ coincide.
To study $P_{_T}^\Lambda(x_{_F},p_{_T})$ in the SIDIS case we then need the full
$\mbox{\boldmath$k$}_\perp$ dependence of the polarizing FF.
To this end we consider a simple gaussian parameterization,
defining

\vspace{-6pt}

\begin{eqnarray}
D_{\Lambda/q}(z,k_\perp) &=&
\frac{d(z)}{M^2}\,\exp\Bigl[\,-\frac{k_\perp^{\!\ 2}}{M^2 f(z)}\,\Bigr]\,,
\label{dkga}\\
\Delta^{\!N}\!D_{\Lambda^\uparrow\!/q}
 (z,\mbox{\boldmath $k$}_\perp) &=&
 \frac{\delta(z)}{M^2}\,\frac{k_\perp}{M}\,
 \exp\Bigl[\,-\frac{k_\perp^{\!\ 2}}{M^2 \varphi(z)}\,\Bigr]\,\sin\phi\,,
\label{dedkga}
\end{eqnarray}       

\noindent
where $\phi$ is the azimuthal angle between the $\Lambda$
intrinsic transverse momentum and the polarization vector.
We use the general relations
$\int d^2k_\perp D_{\Lambda/q}(z,k_\perp) = D_{\Lambda/q}(z)$,
$\int d^2k_\perp\,k_\perp^{\!\ 2}\, D_{\Lambda/q}(z,k_\perp) =
\langle k_\perp^{\!\ 2}(z)\rangle\,D_{\Lambda/q}(z)$.
By imposing the positivity bound
$|\Delta^{\!N}\!D_{\Lambda^\uparrow\!/q}
(z,\mbox{\boldmath$k$}_{\perp})|\ /\ D_{\Lambda/q}(z,k_\perp)
\leq 1$ $\forall\, z\, \mbox{\rm and}\, \mbox{\boldmath$k$}_{\perp}$,
%
%
%
and requiring full consistency with the
approximations and parameterizations adopted in the fitting procedure
to $pp\to\Lambda^\uparrow\,X$ data
[that is, we require that, when
appropriately used into Eq.~(\ref{ptlh}), our parameterizations
(\ref{dkga}), (\ref{dedkga}) obey the simplifying assumption
$\int\! d^2\mbox{\boldmath$k$}_{\perp}\,
F(\mbox{\boldmath$k$}_{\perp})$ $\Rightarrow
F(\langle k_\perp^0\rangle)\,$], we find:

\vspace{-6pt}

\begin{eqnarray}
D_{\Lambda/q}(z,k_\perp) &=&
\frac{D_{\Lambda/q}(z)}{\pi\langle k_\perp^{\!\ 2}(z)\rangle}\,
\exp\left[\,-\frac{k_\perp^{\!\ 2}}
{\langle k_\perp^{\!\ 2}(z)\rangle}\,\right]\,,\\
\mbox{\hspace*{-15pt}}
\Delta^{\!N}\!D_{\Lambda^\uparrow\!/q_v}(z,k_\perp) &\!\!=\!\!&
\Delta^{\!N}\!D_{\Lambda^\uparrow\!/q_v}(z,\langle k_\perp^0\rangle)\,
\frac{4\sqrt{2}}{\sqrt{\pi}}\,
\frac{k_\perp}{\langle k_\perp^{\!\ 2}(z)\rangle^{3/2}}\,
\exp\left[\, -2\,\frac{ k_\perp^{\!\ 2}}
{\langle k_\perp^{\!\ 2}(z)\rangle}
\, \right]\,\!.
\end{eqnarray}

Notice that: 1)~The factor 2 of difference
in the exponential of $\Delta^{\!N}\!D_{\Lambda^\uparrow\!/q_v}$
w.r.t. $D_{\Lambda/q}$ is required by consistency with the
approach in the $p-A$ case, and is by far more stringent than the
most general bounds ; 2)~There is a simple
relation between our ``effective'' $k_\perp^{0}(z)$ and the
physical, observable $\langle k_\perp^{\!\ 2}(z)\rangle$ of the
$\Lambda$ inside the fragmenting jet:
$\langle k_\perp^{\!\ 2}(z)\rangle = 2\,\langle k_\perp^{0}(z)
\rangle^2$. These relations are a very direct consequence of our
approach and can be tested in SIDIS processes.

Finally, the positivity bound reads now

\vspace{-6pt}

\begin{equation}
\frac{|\Delta^{\!N}\!D_{\Lambda^\uparrow\!/q_v}(z,\langle k_\perp^0\rangle)|}
{D_{\Lambda/q}(z)/2} \leq \frac{\sqrt{e}}{2\sqrt{\pi}} \simeq 0.465\,.
\label{pobo2}
\end{equation}

This bound is consistently satisfied by the original
parameterizations obtained for the
$p\,A\to\Lambda^\uparrow\,X$ case.~\cite{abdm01} 

Choosing the $\hat z$-axis along the virtual boson direction,
the $\hat x$-axis along the $\Lambda$ transverse momentum
$\mbox{\boldmath$p$}_T$, the transverse $\uparrow$ direction
results along the positive $\hat y$-axis, and 
$\phi = \pi/2$.
In this configuration, $P_{_T}^\Lambda$ is
given, in the $\ell p\to\ell'\Lambda^\uparrow\,X$ case (HERMES,
H1, ZEUS, COMPASS, E665, etc.) with e.m. contributions only, by
\cite{muta96}

\vspace{-6pt}

\begin{equation}
P_{_T}^\Lambda(x,y,z,p_{_T}) = \frac{\sum_{q}\,e_q^2\,f_{q/p}(x)\,
[\,d\hat\sigma^{\ell q}/dy\,]\,
\Delta^{\!N}\!D_{\Lambda^\uparrow\!/q}(z,p_{_T})}
{\sum_{q}\,e_q^2\,f_{q/p}(x)\,[\,d\hat\sigma^{\ell q}/dy\,]\,
D_{\Lambda/q}(z,p_{_T})}\,.
\label{ptel}
\end{equation}

\noindent

In the case of weak CC processes, $\nu_\mu p\to\mu^-
\Lambda^\uparrow\,X$ (NOMAD, $\nu$-factories, etc.)
one finds ($f_{u/p}(x)=u$, etc.)

\vspace{-6pt}

\begin{equation}
P_{_T}^\Lambda(x,y,z,p_{_T}) =
\frac{(d+R\,s)\Delta^{\!N}\!D_{\Lambda^\uparrow\!/u}+
\bar u\,(\Delta^{\!N}\!D_{\Lambda^\uparrow\!/\bar d}+R\,
\Delta^{\!N}\!D_{\Lambda^\uparrow\!/\bar s})(1-y)^2}
{(d+R\,s)D_{\Lambda/u}+\bar u\,(D_{\Lambda/\bar d}+R\,
D_{\Lambda/\bar s})(1-y)^2}\,,
\label{ptnu}
\end{equation}

\noindent
where $R=\tan^2\theta_c\simeq 0.056$;
notice that at large $x$ and $z$ $P_{_T}^\Lambda\simeq
\Delta^{\!N}\!D_{\Lambda^\uparrow\!/u}\,/\,D_{\Lambda/u}$,
and one may have a direct access to the polarizing FF. 
Analogous expressions hold for the $\bar\Lambda$ case,
by interchanging $D_q$ with $D_{\bar q}$ into (\ref{ptel}),
(\ref{ptnu}), and for the $\bar\nu$ case by interchanging
$q,\,D_q$ with $\bar{q},\,D_{\bar q}$ into (\ref{ptnu}).


As an example, we present in Fig.~1 some preliminary predictions for
$P_{_T}^\Lambda$ and $P_{_T}^{\bar\Lambda}$ vs. $z$ and $p_{_T}$ for
kinematical configurations typical of HERMES and NOMAD experiments.
Our results are compatible with present NOMAD data for
$P_{_T}^{\Lambda}$ in CC interactions;~\cite{nomad00}
however, only few points with large error bars are available. More precise 
data, for different kinematical configurations and at larger energies, are
required for a detailed test of our model and its predictions, and
for more refined parameterizations of the $\Lambda$ polarizing  FF.
We hope that these data will be soon available from running 
or proposed experiments.
 
\begin{figure}[t]
\begin{center}
\hspace*{5pt}
\epsfig{figure=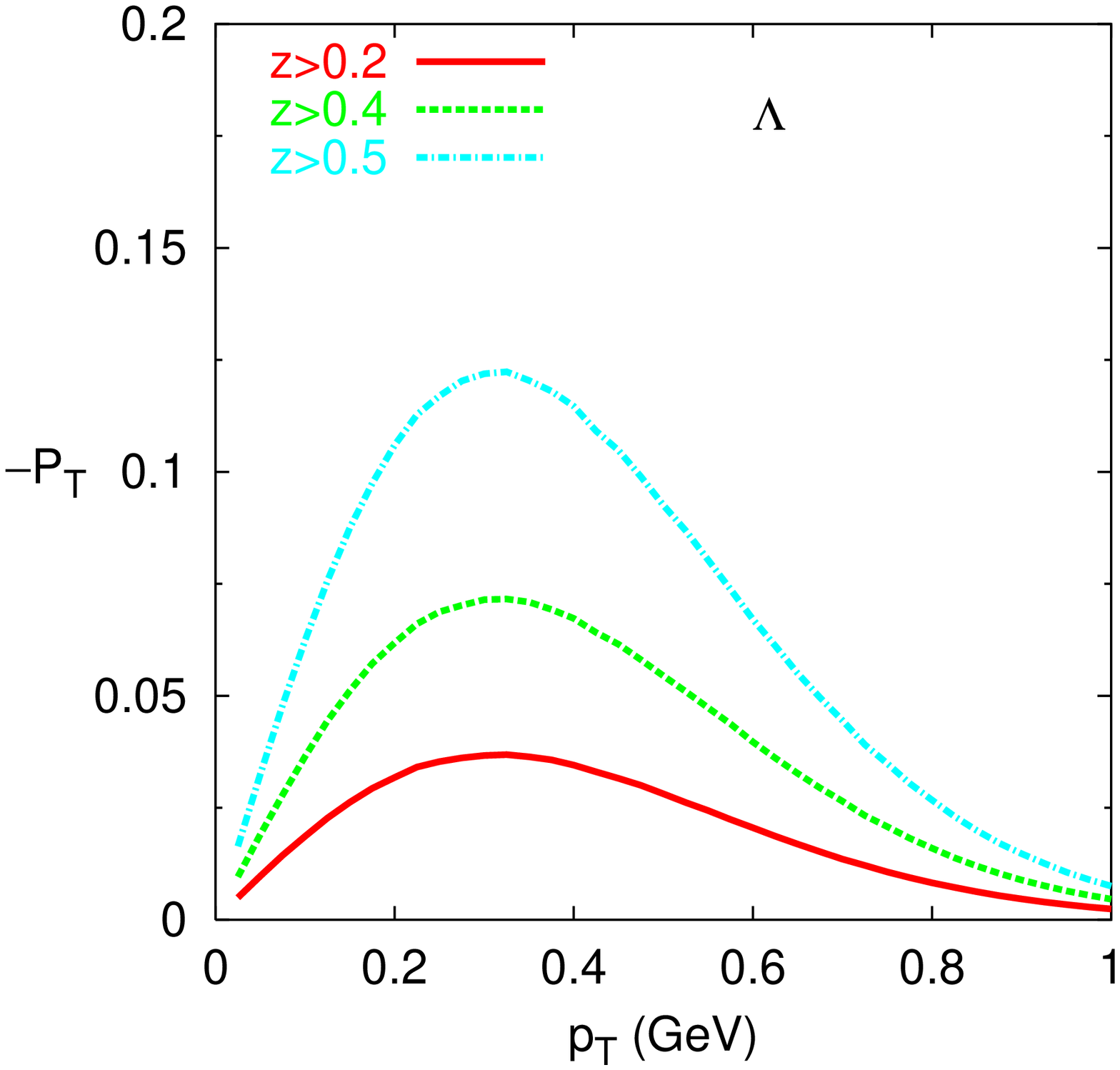,width=5.5truecm,angle=-90}
\hspace*{8pt}
\epsfig{figure=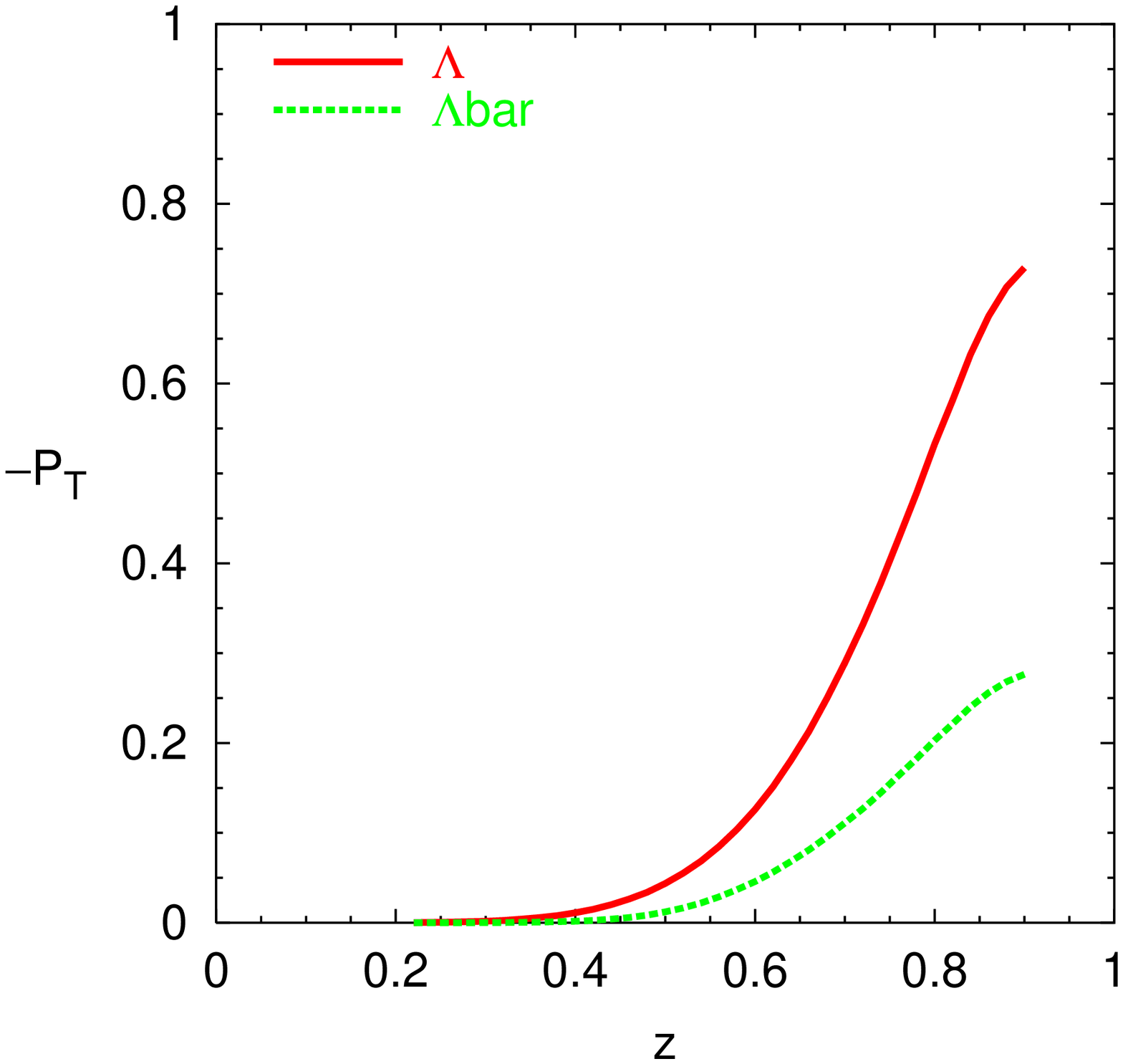,width=5.5truecm,angle=-90}
\end{center}
\vspace*{-4pt}
\caption{Left: $P_{_T}^\Lambda$ vs. $p_{_T}$,
averaged over $z>z_0$ for typical
HERMES kinematics. Right: $P_{_T}^{\Lambda,\bar\Lambda}$ vs. $z$,
averaged over $p_{_T}$ for typical NOMAD kinematics.} 
\vspace{-10pt}
\end{figure}


\end{document}